\documentclass[aps,prb,twocolumn,nofootinbib]{revtex4-1}
\usepackage{epsfig}
\usepackage{graphics}
\usepackage{color}
\usepackage{epstopdf}
\usepackage{amsmath}

\begin{document}
\author{Guo-Hua Liang$^{1}$}
\author{Yong-Long Wang$^{1,2}$}
\author{Hua Jiang$^{2}$}
\author{Meng-Yun Lai$^{1}$}
\author{Hong-Shi Zong$^{1,3,4,}$}~\email[]{Email: zonghs@nju.edu.cn}

\address{$^{1}$Department of Physics, Nanjing University, Nanjing 210093, China}
\address{$^{2}$School of Physics and Electronic Engineering, Linyi University, Linyi 276005, P. R. China}
\address{$^{3}$Joint Center for Particle, Nuclear Physics and Cosmology, Nanjing 210093, China}
\address{$^{4}$State Key Laboratory of Theoretical Physics, Institute of Theoretical Physics, CAS, Beijing, 100190, China}

\title{Spin-polarized transport in helical membranes due to spin-orbit coupling}
\begin{abstract}
The spin-dependent electron transmission through a helical membrane with account of linear spin-orbit interaction has been investigated by numerically solving the Schr\"odinger equation in cylindrical coordinates . It is shown that the spin precession is affected by the magnitude of geometric parameters and chirality of the membrane. This effect is also explained analytically by using the perturbation theory in the weak coupling regime. In the strong coupling regime, the current spin polarization is evident when the number of the open modes in leads  is larger than that of the open channels in the membrane. Moreover, we find that the chirality of the helical membrane can determine the orientation of current spin polarization. Therefore, one may get totally opposite spin currents from the helical membranes rolled against different directions. 

\end{abstract}


\maketitle

\section{INTRODUCTION}
The spintronics and emerging nanoelectronics technology are two crucial fields for the design of next generation nanodevices~\cite{Ahn1754,ko2010ultrathin,ph500144s}. Conventionally, spintronics is divided into two main subfields by metallic and semiconducting materials~\cite{RevModPhys.76.323,sinova2012new,0034-4885-78-10-106001}, where giant magnetoresistance~\cite{PhysRevLett.61.2472}, Aharonov-Casher effect~\cite{PhysRevLett.53.319} and spin-Hall effect~\cite{Kato1910} are discovered. Recently, molecular spintronics~\cite{PhysRevB.92.115418,PhysRevLett.108.218102}, which controls the electron spin transport in organic molecule systems~\cite{caetano2016spin}, has attracted wide attention. This is inspired by the experimental discovery of the high spin selectivity and the length-dependent spin polarization in double-stranded DNA~\cite{Gohler894}, showing the interplay of the molecule structure and spin-orbit coupling (SOC). Since many two-dimensional (2D) nanostructures with complex geometries~\cite{0957-4484-12-4-301,tanda2002crystal,onoe2003structural} have been fabricated successfully, similar situation may also happen in these inorganic materials. Thus the combination of the nanoelectronics and the spintronics provides a platform for exploring the geometric effect on electron spin transport through SOC.

The SOC on some nanostructures has been studied both experimentally~\cite{kuemmeth2008coupling} and theoretically~\cite{PhysRevB.74.155426,jap3452337,PhysRevB.75.085308}, showing geometric influences on band structure and spin polarization. Meanwhile, several theoretical works~\cite{PhysRevB.64.085330,PhysRevB.87.174413,PhysRevB.91.245412,Shikakhwa20161985} have tried to give an effective Hamiltonian with SOC for a general system which is curved and dimensionally reduced. These investigations are based on the thin layer quantization approach~\cite{PhysRevA.23.1982,PhysRevLett.100.230403,Wang201668,PhysRevA.90.042117} in which a confining potential is introduced to constrain particles on a quasi-2D curved surface. Because of the confining potential, the quantum excitation energies in the direction normal to the surface are much greater than those in the tangential directions, then one can safely neglect the quantum dynamics in the normal direction and get an effective 2D Hamiltonian. The treatment is also adopted in our model.

In the present paper, based on the thin layer quantization scheme, we give a brief derivation of the Hamiltonian for a helical membrane with SOC, and investigate the spin polarized transport property accordingly. The Hamiltonian in our model is in fact the same as the case of a tubular two-dimensional electron gases~\cite{PhysRevB.81.075439,PhysRevB.83.115305}, only except that the lateral confining potential or the boundaries are formed by two helices. Recently, spin precession~\cite{PhysRevB.83.115305}, spin polarized current~\cite{PhysRevB.81.075439} and the cross over from weak localization to weak antilocalization~\cite{PhysRevB.93.205306} have been studied for the cylindrical nanowires with SOC, showing the curvature effect on a SOC system. Besides, it is known that the feature of the lateral confining potential in a SOC system could also affect the spin transport prominently. Therefore, we would like to investigate the helical membrane which has both the curvature and boundary effects. Experimentally, this kind of nanostructure can be fabricated in different ways~\cite{Prinz2000828,C4NR00330F} based on different material candidates, such as vapor solid growth for ZnO, strain engineering for InGaAs/GaAs and CVD(Chemical Vapor Deposition) for InGaN.

This paper is organized as follows. In Sec.~\ref{model} we give a brief derivation of the Hamiltonian for a helical membrane with SOC. In Sec.~\ref{Stc} we calculate the spin transport properties by solving the dynamic equation numerically in weak coupling regime, and give an explanation by utilizing the perturbation theory. In Sec.~\ref{Cacp}, we show the spin polarization in the current and analyze the relation between the chirality and features of the spin polarization. We present our conclusions in Sec.~\ref{Conclusion}.


\section{Model}\label{model}

\begin{figure}
  \centering
  \includegraphics[width=0.47\textwidth]{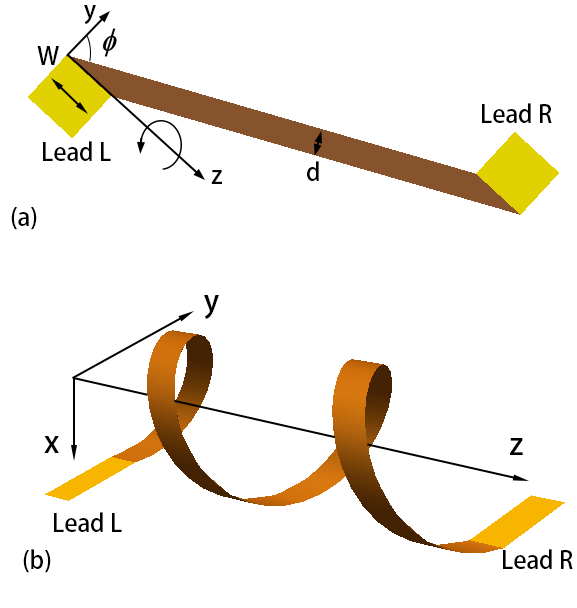}
  \caption{(Color online) (a) Schematic of a two-dimensional membrane with two attached leads in y-z plane. The widthes of the leads and the membrane are $W$ and $d$, respectively. (b) Schematic of a helical membrane with two attached leads. The helical membrane is fabricated by coiling the plane scattering region in (a). The left lead is still in y-z plane.}\label{helm}
\end{figure}

We consider a helical nanomembrane whose edges follow a cylindrical helix around the $z$ axis (see Fig. \ref{helm}(b)). The surface of the membrane is in fact belong to a cylinder, and the helical edges are given by
\begin{equation}
\vec{r}=\rho_0 {\rm cos}\theta \vec{e}_x+\rho_0 {\rm sin}\theta\vec{e}_y+(c \theta +z_0)\vec{e}_z,
\end{equation}
where $\rho_0$ is the radius of the cylinder, $c$ denotes the change of the edges along $z$ direction when the rolling angle $\theta$ increases, and $z_0$ the initial position. For the edges with constant $c$, the total length of the nanomembrane is $L=\theta_0\sqrt{\rho^2+c^2}$, where $\theta_0$ is the total rolling angle.
One can fabricate this kind of nanostructure by scrolling a planar membrane (see Fig. \ref{helm}(a)) which is connected to two leads with width $W$. The longitudinal directions of the lead and the membrane form an angle $\phi={\rm arctan} (c/\rho)$, so the membrane width $d=W {\rm cos}\phi$. To make sure there is no overlap, the parameters should satisfy $2\pi c>W$.

Following da Costa~\cite{PhysRevA.23.1982}, without the SOC, the Hamiltonian for a quantum particle confined on a curved surface is given by
\begin{equation}
H_0=-\frac{\hbar^2}{2 m^*}\frac{1}{\sqrt{g}}\partial_i(\sqrt{g}g^{ij}\partial_j)+V_g, \qquad i,j=1,2,
\end{equation}
where $V_g=-\frac{\hbar^2}{2m^*}(M^2-K)$ is an effective potential induced by the geometry, $m^*$ is the effective electron mass, $M$ and $K$ are the mean and Gaussian curvatures, respectively.

In our case, we assume the spin-orbit field axis is always normal to the surface, namely $\vec{\alpha}=\alpha_3 \vec{e}_3$ (see the Appendix), therefore the effective Hamiltonian of the linear SOC in curvilinear coordinates is
\begin{equation}
H_{so}=-i \frac{\epsilon^{3 a b}}{\sqrt{g}} \alpha_3 \sigma_a \partial_b, \qquad a,b=1,2,
\end{equation}
where $\epsilon^{\mu\nu\lambda}$ is the usual Levi-Civita symbol.

Hence, in the cylindrical coordinate system $(\rho,\theta,z)$,
the entire Hamiltonian reads
\begin{equation} \label{Hami}
H=H_0+H_{so},
\end{equation}
where
\begin{equation}
\begin{aligned}
H_0=-\frac{\hbar^2}{2m^*}\left(\frac{1}{\rho_0^2}\frac{\partial^2}{\partial\theta^2}+\frac{\partial^2}{\partial z^2}
+\frac{1}{4\rho_0^2} \right)+V_\lambda (\theta,z),
\end{aligned}
\end{equation}
\begin{equation} \label{Hso}
\begin{aligned}
H_{so}=-i\alpha_\rho\left[ \right.
(-{\rm sin}\theta \sigma_x+{\rm cos}\theta \sigma_y)\partial _z
-\frac{1}{\rho_0}\sigma_z\partial_\theta \left.\right],
\end{aligned}
\end{equation}
here,$V_\lambda (\theta,z)$ is the transverse confining potential, $\alpha_\rho$ is the SOC strength constant, and $\rho_0$ denotes the curling radius. It is worth noticing that $V_\lambda (\theta,z)$ and $H_{so}(\theta,z)$ together determine the chirality of the system. In the case of reversing $\theta$ direction, $\theta\rightarrow -\theta$, one of $V_\lambda (\theta,z)$ and $H_{so}(\theta,z)$ is reversed, the chirality of the system changes, both of them are reversed, the chirality is invariant.

\section{Spin transport calculation}\label{Stc}
For the study of the spin-polarized transport in the helical nanomembrane, the physical model is that the membrane connected to two leads is coiled around a cylinder, and the inelastic processes take place only in the reservoirs. The two leads with the same width $W$ are planar and tangent to the cylindrical surface. We set that the left lead lies in the y-z plane, the position of the right lead depends on the length of the membrane. The spin-orbit interaction is introduced adiabatically from the leads to the curling region, where the SOC strength constant $\alpha_\rho$ is homogeneous.

The spin-transport problem was solved numerically under the condition of open boundary by using the tight-binding quantum transmitting boundary method (QTBM)~\cite{einspruch2014heterostructures,1.345156}, which is generalized to include the spin degree of freedom in our calculation by representing the on-site and hopping energies by $2\times2$ matrices~\cite{PhysRevB.69.155327}. In the calculation for the SOC Hamiltonian, Hermitian conjugation is used to ensure Hermiticity. We scale the length and the energy in units of $a$ and $t_0=\hbar^2/2m^*a^2$, respectively, where $a=10$nm, and $t_{so}=\alpha_\rho/2a$ is used to measure the strength of the spin-orbit interaction. We employ the critical value $\beta_{so}^c$=$(\pi a/W)^2/(\pi a/W+ak_F)$ defined in planar waveguide to separate the weak and strong coupling regime~\cite{PhysRevB.64.024426}, where $k_F$ is the Fermi wave number. Namely, when $\beta_{so}=t_{so}/t_0<\beta_{so}^c$ the SOC is weak, otherwise, it is strong. We believe that this definition is also applicable to helical membranes approximately. In fact both $\alpha_\rho$ and $W$ determine the contribution of the mixing of the spin subbands, so the increases of $\alpha_\rho$ and $W$ may turn the system from the weak coupling regime to the strong one.

\begin{figure}
  \centering
  \includegraphics[width=0.48\textwidth]{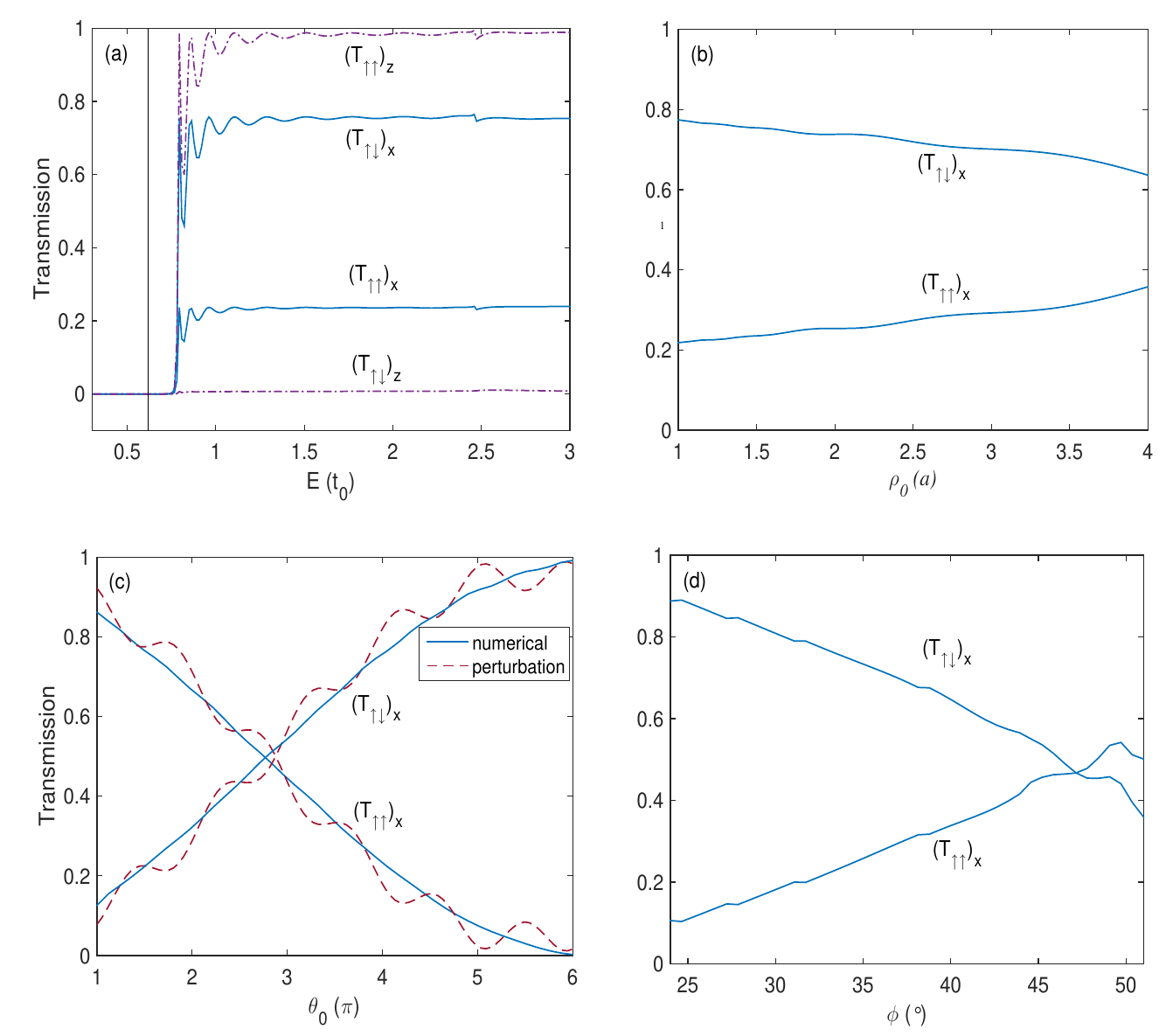}
  \caption{(Color online) Dependence of transmissions through a helical membrane with spin-orbit interaction on (a) energy E, (b) curling radius $\rho_0$, (c) total rolling angle $\theta_0$ (meaning the length of the membrane) and (d) the angle $\phi$. The vertical line in (a) indicates the first threshold energy $\epsilon_1$ in the lead. The dash line in (c) is the analytic result from perturbation theory. $W=4a$ and $t_{so}=0.05$. Other parameters are (a) $r=1.5a$, $\phi=33.7^\circ$, $L=6\pi a$; (b) $E=1.85t_0$, $\phi=33.7^\circ$, $L=6\pi a$; (c) $E=1.85t_0$, $r=1.5a$, $\phi=33.7^\circ$ and (d) $E=1.85t_0$, $r=1.5a$, $L=6\pi a$. }\label{Transmission}
\end{figure}

In this section we consider the weak coupling regime. In Fig. \ref{Transmission} we show the transmissions for helical membranes when purely polarized electrons are injected. Here in $\sigma_z$ representation, the spin-up polarized states with respect to $x,y$ and $z$ axes are expressed as $\frac{\sqrt{2}}{2}(1,1)^T$, $\frac{\sqrt{2}}{2}(1,i)^T$ and $(1,0)^T$, respectively, and  $(T_{ss^\prime})_j$ is used to indicate the probability that an incident electron in a spin-polarized state $s_j$ is scattered into spin state $s_j^\prime$, where $s_j$ denotes the spin orientation along $j$ direction and $s$ is $\uparrow$ or $\downarrow$. As shown in Fig. \ref{Transmission} (a), the step-like behaviour does not happen at the threshold energy $\epsilon_1$ for the ground mode in the lead, but at the energy a little higher. The reason is that the width of the membrane $d$ is smaller than that of the leads $W$, leading to the threshold energy $\tilde{\epsilon}_1$ for the first open channel in the membrane is larger than $\epsilon_1$ in the leads. As described in Fig. \ref{Transmission}, the spin precession effect is evident in $x$ direction ( in $y$ direction, the spin precession is mainly similar to that in $x$ direction, which is not plotted), however there is almost no spin precession for electrons whose initial spins are oriented along $z$ direction. It is also noticed that the spin transmission is almost independent of the incident Fermi energy after some oscillations, which is similar to the case of spin transport in planar quasi-one-dimensional electron gas (Q1DEG) systems~\cite{PhysRevB.64.024426,apl1.102730}. Fig. \ref{Transmission} (b), (c) and (d) show the dependence of spin polarized (in $x$ direction) transmissions on different geometric parameters at the same incident energy, which manifest the geometric effects on the spin precession. For these phenomena, we would like to explain by using perturbation theory.

In the weak coupling regime, the spin-orbit interaction $H_{so}$ can be viewed as a perturbation. The unperturbed Hamiltonian satisfies the eigen equation $H_0|n,s\rangle=E_n^0|n,s\rangle$, where $n$ denotes the subband index and $s$ denotes the spinor. To solve the eigen equation, it is convenient to rewrite the Hamiltonian in a rotated coordinate system $(\tilde{\theta},\tilde{z})$, where
\begin{equation}
\left(
\begin{array}{ccc}
\tilde{\theta} \\ \tilde{z}
\end{array}
\right)=\left(
\begin{array}{ccc}
{\rm cos}\phi & \frac{1}{\rho}{\rm sin}\phi \\
-\rho{\rm sin}\phi & {\rm cos}\phi
\end{array}
\right) \left(
\begin{array}{ccc}
\theta \\ z
\end{array}
\right),
\end{equation}
that is
\begin{equation}
H_0=-\frac{\hbar^2}{2m^*}\left(\frac{1}{\rho^2}\frac{\partial^2}{\partial\tilde{\theta}^2}+\frac{\partial^2}{\partial \tilde{z}^2}
+\frac{1}{4\rho^2} \right)+V^\prime_\lambda (\tilde{z}),
\end{equation}
\begin{equation}
\begin{aligned}
H_{so}=& -i\alpha_\rho \left[ \sigma_\theta(\frac{{\rm sin}\phi}{\rho}
\partial_{\tilde{ \theta}}+{\rm cos}\phi\partial_{\tilde{z}}) \right.\\
& \left. -\sigma_z (\frac{{\rm cos}\phi}{\rho} \partial_{\tilde{\theta}}- {\rm sin}\phi \partial_{\tilde{z}})\right],
\end{aligned}
\end{equation}
where $\sigma_\theta=-{\rm sin}\theta \sigma_x+{\rm cos}\theta \sigma_y$, $V^\prime_\lambda (\tilde{z})$ is the confining potential in $\tilde{z}$ direction, and $\tilde{\theta}$ is the longitudinal direction of the membrane.
We assume the eigen states have the form $|n,s\rangle=e^{i\tilde{k}_n \rho_0\tilde{\theta}} \chi_n(\tilde{z})|s\rangle_z$, where $s=\uparrow$ or $\downarrow$, with the definitions of the spinors $|\uparrow\rangle_z=(1,0)^T$ and $|\downarrow\rangle_z=(0,1)^T$, and $\tilde{k}_n=\sqrt{2m^*(E_n^0-\tilde{\epsilon}_n)}/\hbar$, here $\tilde{\epsilon}_n$ are the solutions of the equation
\begin{equation}
\left[-\frac{\hbar^2}{2m^*}\left(\frac{\partial^2}{\partial\tilde{z}^2}+\frac{1}{4} \right)+V^\prime_\lambda (\tilde{z})\right] \chi_n(\tilde{z})=\tilde{\epsilon}_n\chi_n(\tilde{z}).
\end{equation}
According to degenerate perturbation theory, for each subband $n$ we obtain the following equation
\begin{equation}\label{pert}
\left(
\begin{array}{ccc}
E_n^0+(H_{so})_{nn}^{\uparrow\uparrow}-E & (H_{so})_{nn}^{\uparrow\downarrow} \\
(H_{so})_{nn}^{\downarrow\uparrow} & E_n^0+(H_{so})_{nn}^{\downarrow\downarrow}-E
\end{array}
\right) \left(
\begin{array}{ccc}
a_{n\uparrow}^0\\
a_{n\downarrow}^0
\end{array}
\right)=0,
\end{equation}
where $a_{ns}^0$ are the zeroth order coefficients used to expand the perturbed states in terms of the unperturbed states $|n,s\rangle$, and $(H_{so})_{nn}^{ss^\prime}=\langle n,s| H_{so} |n, s^\prime\rangle$ are the matrix elements expanded in the subspace of each subband $n$. Due to the reflection symmetry of the transverse confining potential $V_\lambda ^\prime(\tilde{z})$ in $\tilde{z}$ direction, the term $\langle n| \partial_{\tilde{z}}|n\rangle $ vanishes.
It is straight forward to obtain the eigenvalues of Eq. (\ref{pert})as
\begin{equation}\label{dispe}
E^{\pm}=\tilde{\epsilon}_n+\frac{\hbar^2 \tilde{k}_n^2}{2m^*}\pm \alpha_\rho \tilde{k}_n.
\end{equation}
This dispersion relation for the helical membrane is the same as in the planar case in the weak SOC regime.
The energy splitting means that electrons with the same energy may have different wave vectors, that is $E^+(\tilde{k}_n^+)=E^-(\tilde{k}_n^-)$, leading to $\tilde{k}_n^--\tilde{k}_n^+=2m^*\alpha_\rho/\hbar^2$. As this difference only depends on the SOC strength constant, the transmissions in Fig.~\ref{Transmission}(a) show energy independent behaviors.
The corresponding eigenvectors without normalization are $[1-{\rm cos}\phi,i e^{i \theta}{\rm sin}\phi]^T$ and $[i e^{-i\theta}{\rm sin}\phi, 1-{\rm cos}\phi]^T$. With the consideration of initial condition, if we were to driven $|\uparrow\rangle_z$ polarized electrons into the system, the wave emerging from the conductor will be represented as
\begin{equation}
\begin{aligned}
\psi^{\uparrow_z}=& \left[
\begin{array}{ccc}
{\rm cos}^2(\phi/2)e^{i\beta}+{\rm sin}^2(\phi/2)e^{-i\theta} \\ \frac{i}{2} {\rm sin}\phi (1-e^{i\theta+i\beta })
\end{array}
\right] e^{i \tilde{k}_n^+ L}
\end{aligned}
\end{equation}
where $L$ is the length of the helical membrane, and $\beta=2m^*\alpha_\rho L/\hbar^2$. While if the incident electrons are in $|\uparrow\rangle_x$ state or $|\uparrow\rangle_y$ state, the outgoing wave will be represented as
\begin{equation}
\begin{aligned}
\psi^{\uparrow_x}=& \frac{\sqrt{2}}{4}
\left[
\begin{array}{ccc}
e^{-i\theta}(1-e^{i\phi})+(1+e^{i\phi})e^{i\beta} \\ 1+e^{i\phi}+(1-e^{i\phi})e^{i\theta+i\beta}
\end{array}
\right] e^{i \tilde{k}_n^+ L},
\end{aligned}
\end{equation}
or
\begin{equation}
\begin{aligned}
\psi^{\uparrow_y}=& \frac{\sqrt{2}}{4} \left\{
\left[
\begin{array}{ccc}
1+{\rm cos}\phi-{\rm sin}\phi \\ (1-{\rm sin}\phi-{\rm cos}\phi)(ie^{i\theta})
\end{array}
\right]e^{i \tilde{k}_n^- L} \right. \\
&+ \left. \left[
\begin{array}{ccc}
(1-{\rm cos}\phi+{\rm sin}\phi)e^{-i\theta} \\ i(1+{\rm cos}\phi+{\rm sin}\phi)
\end{array}
\right]e^{i \tilde{k}_n^+ L} \right\}.
\end{aligned}
\end{equation}
And then the corresponding transmission probabilities are
\begin{equation}\label{Tz}
(T_{\uparrow\uparrow})_z=1-{\rm sin}^2\phi {\rm sin}^2\frac{\beta+\theta_0}{2},
\end{equation}
\begin{equation}\label{Tx}
\begin{aligned}
(T_{\uparrow\uparrow})_x=\frac{1}{2}\left[1+{\rm cos}\beta {\rm cos}^2\frac{\phi}{2}+{\rm cos}(\beta+2\theta_0){\rm sin}^2\frac{\phi}{2}\right]
\end{aligned}
\end{equation}
and
\begin{equation}
\begin{aligned}
(T_{\uparrow\uparrow})_y=& \frac{1}{2}\left[1+{\rm sin}^2\phi {\rm cos}\theta_0
+{\rm cos}^2\phi {\rm cos}\theta_0 {\rm cos}(\beta+\theta_0)  \right. \\
& \left. +{\rm cos}\phi{\rm sin}\theta_0 {\rm sin}(\beta+\theta_0) \right] .
\end{aligned}
\end{equation}
In this method, the reflectivity is neglected, so $T_{\uparrow\downarrow}=1-T_{\uparrow\uparrow}$. Thus far, it is clear that the transmissions are determined by the geometry of the helical membrane through the three dimensionless parameters, namely $\beta, \theta_0$ and $\phi$. From Eq.~\eqref{Tz}, it is easy to see $(T_{\uparrow\downarrow})_z<{\rm sin}^2\phi$, that is, for small $\phi$, the spin-polarization in $z$ direction is maintained during the transport, as shown in Fig.~\ref{Transmission}(a). In the limit of $\phi \rightarrow 0$, the results above show that there is no spin precession in $z$ (transverse) direction, and the spin polarized transmissions for $x$ (the normal) direction and $y$ (the longitudinal) direction are $(T_{\uparrow\uparrow})_x=(T_{\uparrow\uparrow})_y={\rm cos}^2 (\beta/2)$, similar to the planar case in weak SOC regime. In Fig.~\ref{Transmission} the angle $\phi$ can not be too big (otherwise the width $d$ will be too narrow for transport), therefore we have $(T_{\uparrow\uparrow})_x \approx (T_{\uparrow\uparrow})_y$. If we only change the radius $\rho_0$ and keep other parameters fixed, then in Eq.~\eqref{Tx} only $\theta_0$ is changed, which means that the change of the transmission can not exceed $\frac{1}{2}{\rm sin}^2\frac{\phi}{2}$, thus in Fig.~\ref{Transmission}(b) the transmission is not considerably sensitive to the radius. Fig.~\ref{Transmission}(c) shows that the analytic result Eq.~\eqref{Tx} is broadly in line with the numerical result except some oscillations. Note that, comparing with the numerical calculation, we ignore the effects of leads and their connections to the membrane in the perturbation theory.

\begin{figure}
  \centering
  \includegraphics[width=0.49\textwidth]{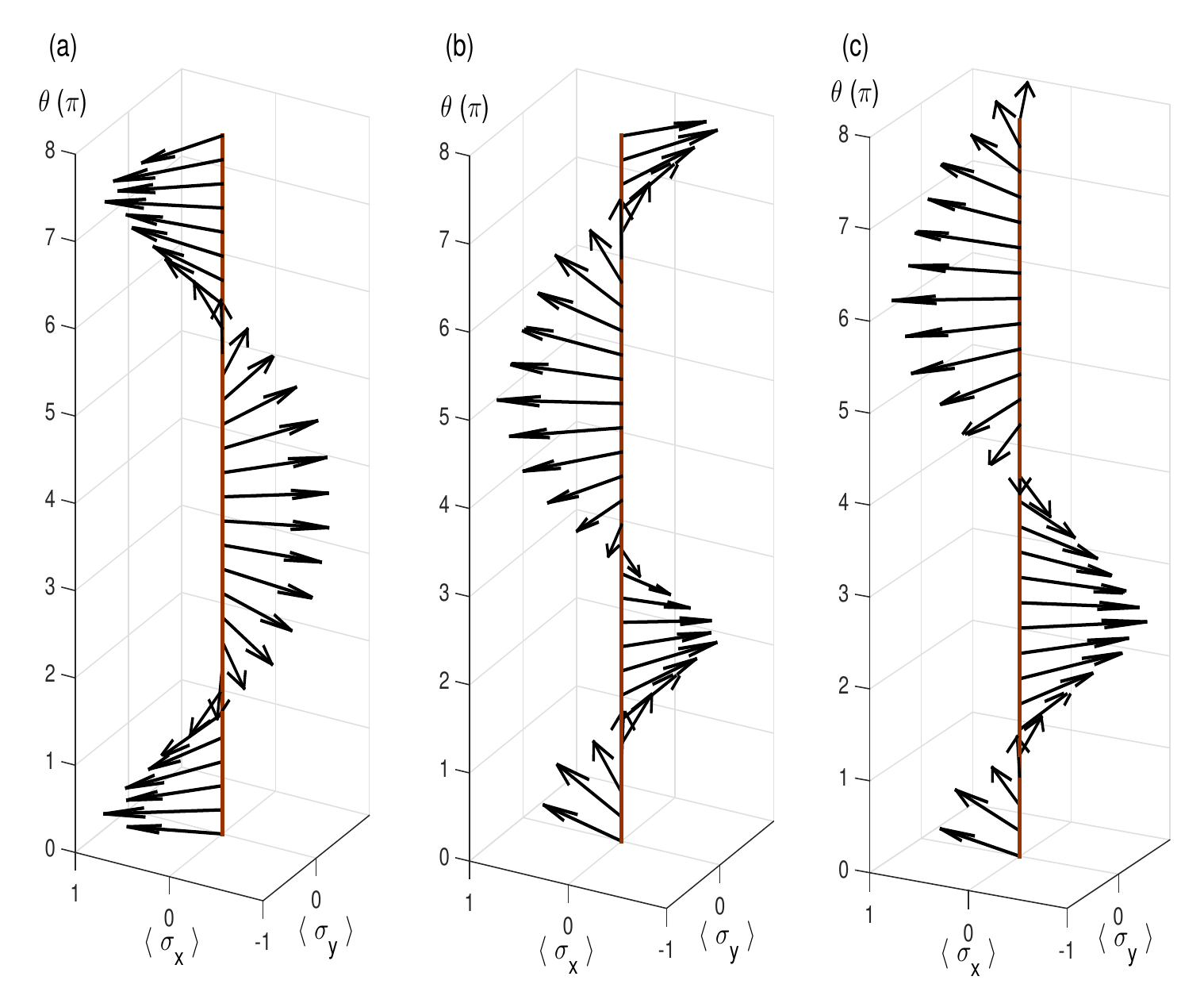}
  \caption{(Color online) Spin precession of electrons in the middle line of (a) the left-handed helical membrane, (b) the planar membrane and (c) the right-handed helical membrane, respectively. Here $W=4a$, $E=1.85t_0$, $\phi=30.5^\circ$, $t_{so}=0.1$, $L=8\pi\sqrt{\rho_0^2+c^2} $, for (a) and (c), $\rho_0=1.5a$.}\label{polar}
\end{figure}

From the wavefunctions above, we find that the coiling can affect the spin precession along $z$ axis significantly. In the case of the initial spin orientation along the $x$ direction, we investigate the spin precession in the middle of the membrane, the results are described in Fig.~\ref{polar}. It is found that at the same injection energy and SOC constant, the spin precessions present different characteristics for the membranes with different chirality. The spin precesses counterclockwise in the left-handed helical membrane, while for the right-handed one, the precession is clockwise. We also depict the situation in a planar membrane (see Fig.~\ref{polar}(b)), where the direction of spin precession (here we show a clockwise case) in fact depends on the orientation of the spin-orbit field axis (parallel or antiparallel to the normal direction of the surface). However, no matter it's counterclockwise or clockwise, the period of the precession in planar case is different from that in the helical one.

\section{Chirality and spin current}\label{Cacp}
\begin{figure}
  \centering
  \includegraphics[width=0.48\textwidth,height=0.54\textwidth]{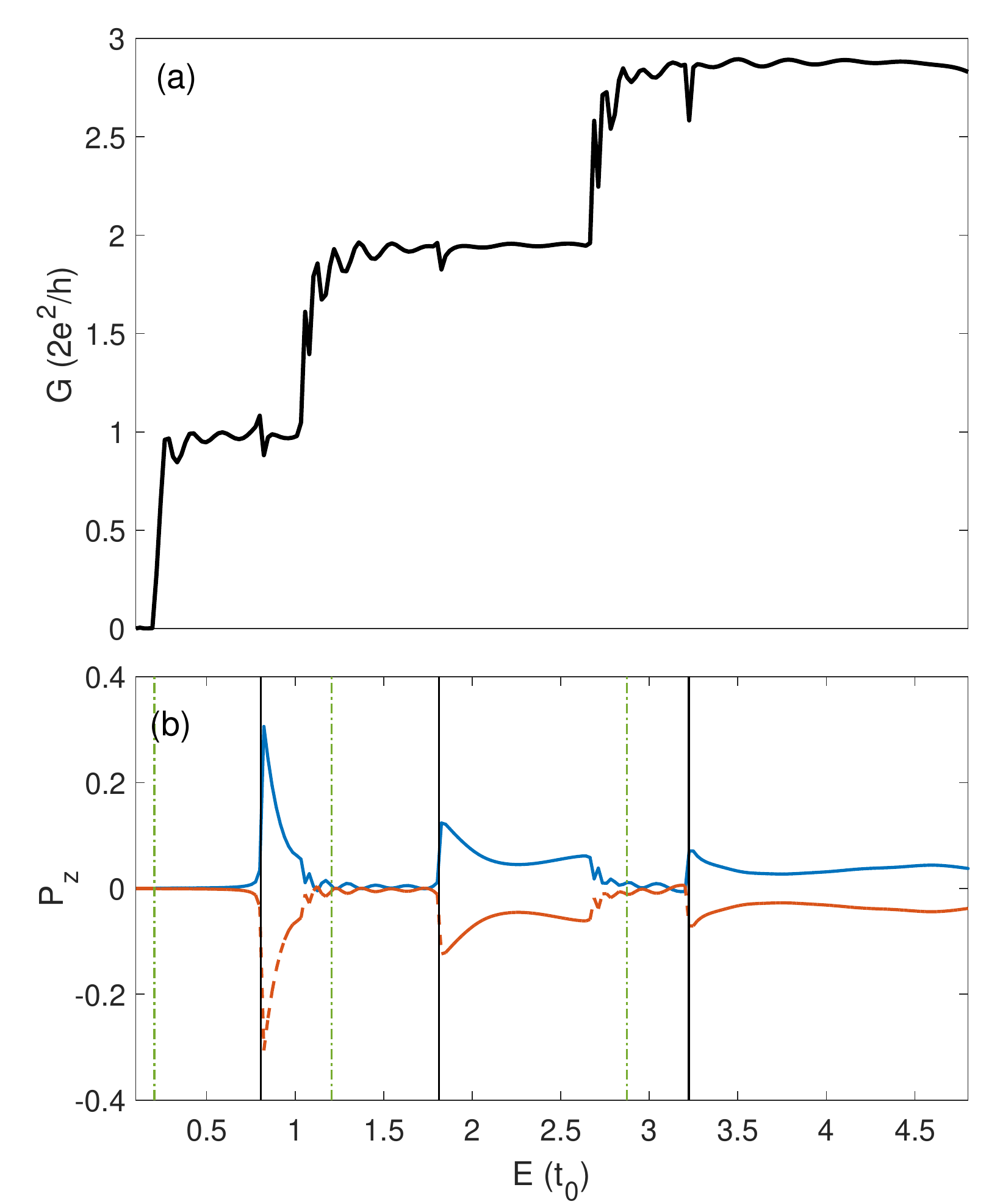}
  \caption{(Color online) (a) Conductance versus injection energy for the helical membrane with opposite chirality. (b) Mean spin polarization $P_z$ of the current at the right lead in the presence of spin-orbit interaction. The solid and dashed curves are for right handed and left handed helical membranes, respectively. The solid vertical lines represent $\epsilon_2$, $\epsilon_3$ and $\epsilon_4$, and dashed-dotted vertical lines represent $\tilde{\epsilon}_1$, $\tilde{\epsilon}_2$ and $\tilde{\epsilon}_3$. The parameters are $W=7a$, $\rho_0=1.4a$, $\theta_0=4\pi$, $t_{so}=0.3$, $\phi=39^\circ$.}\label{CP}
\end{figure}

The spin polarization of the outgoing current is the ratio between normalized spin conductance and the total conductance. The definition is given by
\begin{equation}
P_j=\sum_{s_j} \frac{G_{s_j,s_j^\prime}-G_{s_j,-s_j^\prime}}{G}, \qquad j=x,y,z.
\end{equation}
It has been proved that in an arbitrary structure which attaches to two leads, the spin polarization does not appear if the incident electrons are in the first energy subband~\cite{PhysRevB.66.075331,PhysRevLett.94.246601}. Thus to generate spin polarization current in our model, intersubband transmission is necessary, which means strong SOC should be considered. In this situation, based on Landauer formula~\cite{Landauer1987}, the conductance and the spin polarization $P_z$ of the current are plotted in Fig.~\ref{CP} for the helical membrane with opposite chirality.  The conductance shows a step-like dependence on the incident energy, corresponding to the appearance of new open channels in the membrane. We have mentioned that the threshold energy $\epsilon_n$ for the $n$th mode in the leads is always smaller than the energy $\tilde{\epsilon}_n$ for the $n$th channel in the membrane, which is due to the difference between their widths. Therefore even the increasing energy opens new modes in the leads, the conductance does not get a new step until the new channel is opened in the coiling region. We also find that reversing the chirality doesn't change the charge conductance.

For the spin polarization $P_z$ of the current, we observe that the spin polarization is obvious when the energy satisfies $\epsilon_n<E<\tilde{\epsilon}_n$. In this energy range, the number of open modes in the leads is always greater than the number of open channels in the membrane, leading to an abrupt change of intersubband mixing. While this situation does not happen for $E<\epsilon_2$, because that there exists only one open mode in the leads. For any two-terminal structures, no spin polarization can occur when the leads support only one open mode. This conclusion has been proven by the calculation of transmissions~\cite{PhysRevB.66.075331} and the analysis of the symmetries of S matrix~\cite{PhysRevLett.94.246601}. For each energy range with $n>2$, the polarization peak decreases slowly. The reason of the decreasing peak is that, at the beginning of the energy range, a new mode $n$ is opened in the lead, while because of the subband splitting due to SOC in the channel, the state $|n,\uparrow\rangle$ can be transmitted much easier to the right lead than $|n,\downarrow\rangle$. With the incident energy increasing, the state $|n,\downarrow\rangle$ participates in the transmission more and more, which leads to the cancellation of the positive spin polarization of the current to some extent.

\begin{figure}
  \centering
  \includegraphics[width=0.48\textwidth]{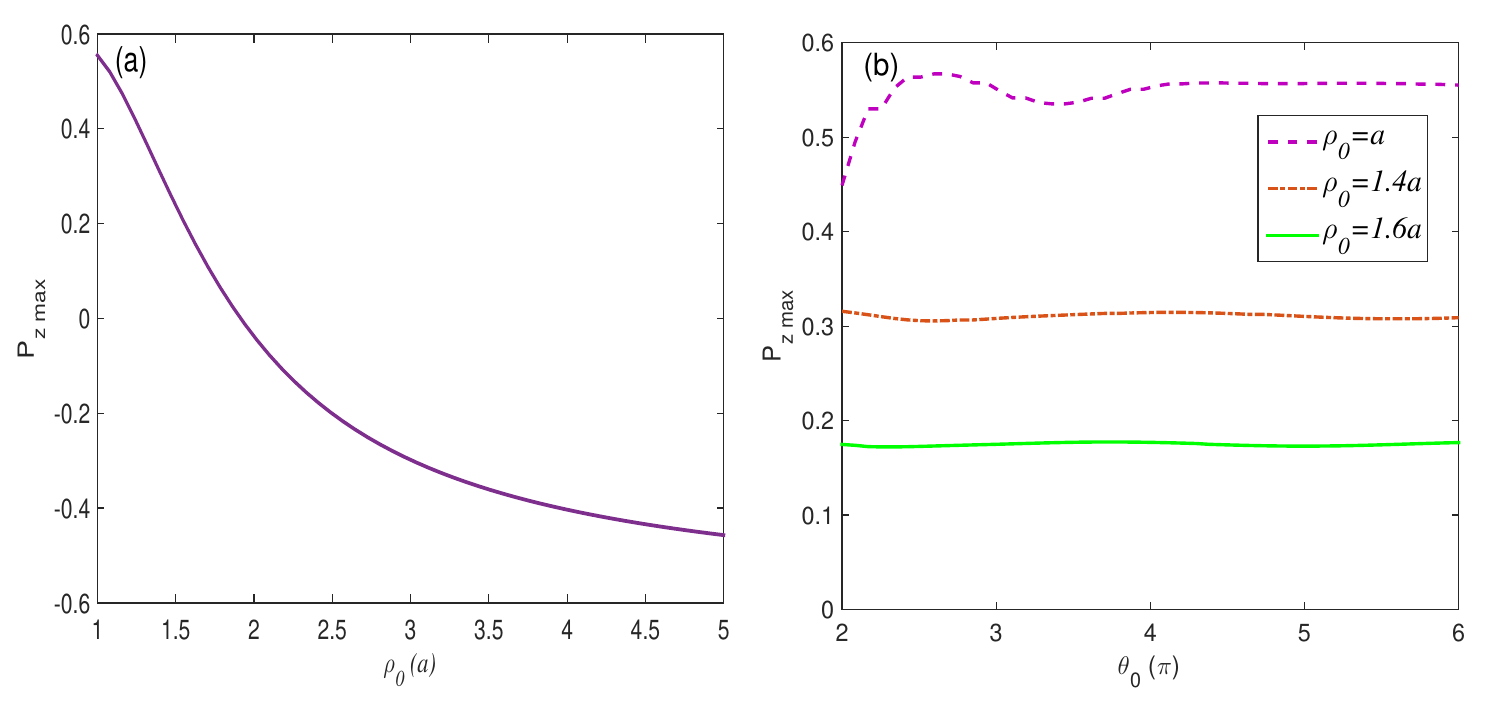}
  \caption{(Color online) The peak values of the current spin polarization at the right lead of right-handed helical membranes, as a function of (a) the coiling radius $\rho_0$ and (b) the channel length.
   }\label{Prho}
\end{figure}

We furthermore find that the spin polarization is obviously affected by the radius $\rho_0$. In Fig.~\ref{Prho} we plot the dependence of peak values of the current spin polarization $P_{z \max}$ in the energy range $\epsilon_2<E<\tilde{\epsilon}_2$, on the curvature radius $\rho_0$ and on the channel length indicated by $\theta_0$ (Note that $\rho_0=a$ in fact violates the expression $2\pi c>W$ which insures no overlapping portion in the membrane). It shows that the curvature can evidently affect the magnitude and sign of the spin polarization of the current. With increasing the radius, the curvature effect becomes weak and the change of the spin polarization becomes gentler. The cause is that the lifting of spin degeneracy introduced by the SOC Hamiltonian~\eqref{Hso} is influenced by the curling radius. We note that the peak decreases and even vanishes when $\rho_0\approx 2a$, then increases with an opposite sign, which is due to the shift of spin subbands with the curvature.
Moreover, when the radius decreases, the lower geometric potential also has weak impact on the energy subband, leading to a little change of the spin polarization. Fig.~\ref{Prho}(b) shows that the channel length has little impact on the peak values. This is to be expected since the peak is determined by the intrasubband and intersubband transmissions between the same spin states(which will be explained below), although the channel length could affect the energy levels of the membrane through the geometric potential $V_g$, however, here the geometric potential is so weak that this effect is negligible. For the line of $\rho=a$ the peak values change with the channel length at the beginning, this may be due to the shortness of the channel which makes the two scattering regions too close to form a stable travelling wave between them.

\begin{figure}
  \centering
  \includegraphics[width=0.49\textwidth]{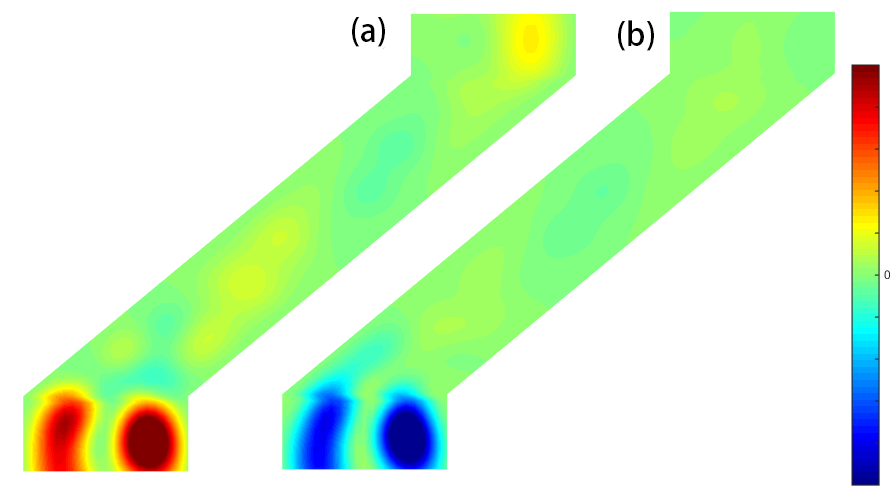}
  \caption{(Color online)
  Spin densities for transport through a right-handed helical membrane when injected electrons are in $n=2$ state, $E=0.9t_0$. For (a) and (b), the incident electrons are spin-up and spin-down polarized, respectively. }\label{zspin}
\end{figure}

In our model the Hamiltonian~\eqref{Hami} is invariant under time reversal, that is $[T,H]=0$, where the time-reversal operator $T=i\sigma_y C$, with $C$ is complex-conjugation operator. Likewise, as we mentioned, the total Hamiltonian is also invariant under the reflection transformation with respect to $\theta$ direction. The two symmetries together result in~\cite{PhysRevLett.94.246601}
\begin{equation}
T_{n_i s_i,n_o s_o}=T_{n_o s_o, n_i s_i},
\end{equation}
where $n_i$ and $n_o$ denote the subband indexes in the incoming and outgoing leads, respectively, and the same meaning for the subscripts of spin states. Considering the definition of spin conductance, we can conclude that $G_{\uparrow\downarrow}=G_{\downarrow\uparrow}$, leading to
\begin{equation}\label{Pj}
P_j=\frac{G_{\uparrow_j,\uparrow_j}-G_{\downarrow_j,\downarrow_j}}{G}.
\end{equation}
This indicates that the transmissions between opposite spin states have no contribution to the spin polarization of the outgoing current since they cancel each other out. Hence the nonzero polarization is caused by the transmissions between the same spin states, which includes intrasubband and intersubband contributions.
To visually comprehend the generation of the spin polarization, the spin density $\langle \sigma_z \rangle$ is plotted in Fig.~\ref{zspin}(a) and (b) for two situations: all the incident electrons are in state $|2,\uparrow\rangle$ or $|2,\downarrow\rangle$. At the energy $E=0.9t_0$, there are only two open modes in the leads and one open channel in the membrane. It is found that some of the electrons with spins orienting along the $z$-direction are able to be transmitted to the outgoing lead, while the spin negative mode is almost reflected completely. This difference demonstrates the asymmetry of transmissions for opposite spin states, which leads to outstanding spin polarization in the current. We also note that in the channel the similar oscillation with same sign appears for the two cases. For both cases the incident waves are reflected highly, which indicates that the connection between the left lead and the channel acts as a closed boundary approximately. Considering the continuity condition, we could conclude that in both cases the wavefunctions in the channel are similar since they have similar boundary conditions. Hence, it looks like the oscillation is independent of the spin orientation of the incident wave.

In addition, we find that the polarization $P_z$ shows entirely opposite character for the helical membrane with different chirality. This can be explained by analyzing symmetries of spin transport in these two different structures. In our model the SOC Hamiltonian ~\eqref{Hso} is invariant under the operation $K=R_\theta (SR_x) (SR_z)$, where $ R_\theta(\theta)=-\theta$ and $SR_j(\sigma_j)=-\sigma_j$ reversing the spin quantum axis in $j$ direction. The Hamiltonians for the two oppositely coiled helical membranes have the relation $H_{so}^R=K (SR_x) (SR_z) H_{so}^L $, where the superscript $R$ and $L$ denote the right-handed and left-handed helical membranes, respectively. This implies that
\begin{equation}
|s\rangle_l^R=|-s\rangle_l^L, \qquad l=x,z.
\end{equation}
From the definition Eq.~\eqref{Pj}, it is straightforward to obtain $P_l^R=-P_l^L$. Using this property, one might be able to generate and cancel the spin polarization by adding this kind of nanostructures in a circuit. As the conduction electrons go out from the planar right lead, we expect the developing techniques of spin-polarization detection for planar 2D semiconductor could demonstrate this chirality dependent property experimentally. One of the possible ways is using the all-electrical structures~\cite{nphys543,PhysRevB.79.165321} which is based on the fact that the spin polarization results in an electrochemical potential difference, leading to a change of the nonlocal voltage between a magnetic contact and a nonmagnetic contact.

Moreover, under the transformation $K^\prime=R_{\theta} C$, the SOC Hamiltonian ~\eqref{Hso} is also invariant. Therefore, we have $H_{so}^R=K^\prime C H_{so}^L $, which means the eigenfunctions in the helical membrane with opposite chirality have the relation
\begin{equation}
\left(
\begin{array}{ccc}
\psi_1^R \\ \psi_2^R
\end{array}
\right)_n=\left(
\begin{array}{ccc}
\psi_1^{L *} \\ \psi_2^{L *}
\end{array}
\right)_n.
\end{equation}
So for a helical membrane with closed boundary, considering the definition of spin density $\langle \sigma_y \rangle = i (\psi_1^* \psi_2-\psi_2^*\psi_1)$, we obtain $\langle \sigma_y \rangle ^R=-\langle \sigma_y \rangle^L$, manifesting that the chirality of the helical membrane could reverse the spin orientation along $y$ direction.

\section{Conclusion}\label{Conclusion}
In summary, we have showed that the spin precession in a helical membrane with spin-orbit interaction is sensitive to the geometry of the system in both analytic and numerical calculations. The longitudinal spin precession about the cylindrical axis manifests clockwise or couterclockwise features for different chirality. In the strong coupling regime, we demonstrate that intersubband mixing induces a spontaneous spin polarization of the outgoing current, besides, this polarization becomes prominent when the injection energy is in the region where the number of the open modes in the leads is larger than that of the open channels in the membrane. In this case, changing the coiling radius could enhance, weaken or even change the direction of the spin polarization.
Further, for the helical membrane with inverse chirality, the spin polarization of the current reveals the same magnitude but opposite direction. This provides a possible way to control the spin polarization through the geometry of the SOC system.

\acknowledgments

This work is supported in part by the National Natural Science Foundation of China (under Grants No. 11275097, No. 11475085, No. 11535005 and No. 11690030).

\appendix
\section{Spin-orbit interaction on a curved surface}
The linear spin-orbit interaction Hamiltonian for electrons can be generalized to a curved three-dimensional manifold~\cite{PhysRevB.91.245412} as follows
\begin{equation}
H_{so}=-i \frac{\epsilon^{\mu \nu \lambda}}{\sqrt{G}} \alpha_\mu \sigma_\nu \partial_\lambda, \qquad \mu,\nu, \lambda=1,2,3,
\end{equation}
where $\epsilon^{\mu\nu\lambda}$ is the usual Levi-Civita symbol, $G$ is the determinant of the metric tensor of the manifold, $\alpha_\mu$ is the spin-orbit interaction constant, $\sigma_\nu$ is the generator of the Clifford algebra in curved space $\{\sigma_\mu,\sigma_\nu\}=2G_{\mu\nu}$.

In curvilinear coordinates $(q^1,q^2,q^3)$, we consider a surface $S$ which is parametrized by $\vec{r}=\vec{r}(q^1,q^2)$, thus the three-dimensional space in the immediate neighbourhood of $S$ can be parametrized as $\vec{R}=\vec{r}(q^1,q^2)+q^3\vec{n}(q^1,q^2)$, where $\vec{n}$ is the unit vector normal to $S$. The derivatives of $\vec{r}$ and $\vec{n}$ satisfy the relation
\begin{equation}
\frac{\partial \vec{n}}{\partial q^a}=\alpha_a^b \frac{\partial \vec{r}}{\partial q^b}, \qquad a,b=1,2,
\end{equation}
where $\alpha_a^b$ are determined by the Weingarten equations. According to this relation, we obtain
\begin{equation}
G=f^2 g,
\end{equation}
where $f=1+q^3 {\rm Tr}(\alpha_a^b)+(q^3)^2 {\rm det}(\alpha_a^b)$, $G$ and $g$ are the determinants of the two metric tensors $G_{\mu\nu}=\frac{\partial \vec{R}}{\partial q^\mu} \cdot \frac{\partial \vec{R}}{\partial q^\nu}$ and  $g_{ab}=\frac{\partial \vec{r}}{\partial q^a} \cdot \frac{\partial \vec{r}}{\partial q^b}$, respectively.

In the thin-layer quantization procedure for the Hamiltonian without SOC, a rescaled wavefunction $\chi=\sqrt{f}\psi$ is introduced, so the transformation should also be applied to the eigenequation $H_{so}\psi=E_{so}\psi$, that is
\begin{equation}
\begin{aligned}
E_{so}\chi=\sqrt{f}\left(-i \frac{\epsilon^{\mu \nu \lambda}}{\sqrt{G}} \alpha_\mu \sigma_\nu \partial_\lambda \right)\frac{\chi}{\sqrt{f}}.
\end{aligned}
\end{equation}
In the $q^3\rightarrow 0$ limit, at the zeroth order we obtain
\begin{equation}\label{eqso}
E_{so}\chi=-i \left[\frac{\epsilon^{\mu \nu a}}{\sqrt{g}} \alpha_\mu \sigma_\nu \partial_a \chi +\frac{\epsilon^{a b 3}}{\sqrt{g}} \alpha_a \sigma_b (\partial_3-M) \chi \right],
\end{equation}
where $\mu,\nu=1,2,3$ and $a,b=1,2$, $M={\rm Tr}(\alpha_a^b)/2$ is the mean curvature. The term containing $\partial_3\chi$ prevents the separability of the quantum dynamics along the tangential direction of the surface from the normal quantum motion when the coupling constants $\alpha_1$ and $\alpha_2$ do not vanish. However, if the confining potential normal to the surface is large enough, this term can be treated perturbatively, then we get $\partial_3 \chi |_{q^3\rightarrow 0}\approx 0$. The third term is a geometric spin-orbit field induced by the mean curvature and SOC in the surface. Generally, the coupling constant vector $\vec{\alpha}$ is determined by the direction and the magnitude of electrostatic field~\cite{PhysRevLett.108.218102}, $\vec{\alpha}=\frac{\hbar^2}{4 m^2 c^2}\vec{\nabla} V$, with the crystal potential $V$, the electron mass $m$, the speed of light $c$. If $\vec{\nabla}V$ always points to the direction normal to the surface, $\alpha_1$ and $\alpha_2$ vanish, then only the first term in Eq. (\ref{eqso}) exists in the effective SOC Hamiltonian, and the quantum motions in the tangential directions and the normal direction of the surface can be separated perfectly. This is also the case in our model.



\end{document}